\documentclass[twocolumn,showpacs,preprintnumbers,amsmath,amssymb]{revtex4}
\usepackage{dcolumn}
\usepackage{bm}
\usepackage{graphicx}
\DeclareGraphicsExtensions{.jpg,.pdf,.mps,.png,.eps,.ps,.EPS}

\begin{document}
\title{Topological Generalizations of network motifs}

\author{N. Kashtan$^{1,2}$, S. Itzkovitz$^{1,3}$, R. Milo$^{1,3}$, U. Alon$^{1,3}$}

\affiliation{ $^1$Department of Molecular Cell Biology\\
$^2$Department of Computer Science and Applied Mathematics\\
$^3$Department of Physics of Complex Systems, Weizmann Institute
of Science, Rehovot, Israel 76100\\
}

\begin{abstract}
Biological and technological networks contain patterns, termed
network motifs, which occur far more often than in randomized
networks. Network motifs were suggested to be elementary building
blocks that carry out key functions in the network. It is of
interest to understand how network motifs combine to form larger
structures. To address this, we present a systematic approach to
define 'motif generalizations': families of motifs of different
sizes that share a common architectural theme. To define motif
generalizations, we first define 'roles' in a subgraph according
to structural equivalence. For example, the feedforward loop
triad, a motif in transcription, neuronal and some electronic
networks, has three roles, an input node, an output node and an
internal node. The roles are used to define possible
generalizations of the motif. The feedforward loop can have three
simple generalizations, based on replicating each of the three
roles and their connections. We present algorithms for efficiently
detecting motif generalizations. We find that the transcription
networks of bacteria and yeast display only one of the three
generalizations, the multi-output feedforward generalization. In
contrast, the neuronal network of \emph{C. elegans} mainly
displays the multi-input generalization. Forward-logic electronic
circuits display a multi-input, multi-output hybrid. Thus,
networks which share a common motif can have very different
generalizations of that motif. Using mathematical modelling, we
describe the information processing functions of the different
motif generalizations in transcription, neuronal and electronic
networks.

\end{abstract}

 \maketitle

\section{Introduction}

A major current challenge is to understand the function of
biological information-processing networks
~\cite{Hartwell,Ouzounis,McAdams2000,Elowitz,Savageau,Rao,Strogatz,Bolouri,Hasty,Guet,Tyson,Sneppen,NewmanReview}.
These networks, as well as networks from engineering, ecology, and
other fields, were recently found to contain \emph{network
motifs}: small subgraphs that occur in the network far more often
than in randomized networks ~\cite{Milo,Shen-Orr}. Each class of
networks was found to have a characteristic set of network motifs
~\cite{Milo_2004}. Information processing networks, such as gene
regulation networks ~\cite{Shen-Orr,Lee}, neuron networks, and
some electronic circuits, were found to share many of the same
network motifs ~\cite{Milo,Milo_2004}. Recently, in the case of
the transcription network of the bacterium \emph{E. coli}, network
motifs were shown theoretically and experimentally to function as
elementary building blocks of the network, each performing
specific information-processing tasks
~\cite{Shen-Orr,Mangan,Mangan_JMB}. For example, one of the most
significant motifs shared by biological information processing
networks is the feedforward loop (FFL). In transcription networks,
the feedforward loop with positive regulations was shown to act as
a 'persistence detector' circuit that rejects transient activation
signals yet allows rapid response to inactivation signals
~\cite{Shen-Orr, Mangan,Mangan_JMB}. A second motif, the
single-input module, was shown to generate a temporal order of
gene expression, which correlates with the functional order of the
genes in the pathway ~\cite{Ronen,Zaslaver}. A third major motif,
the bifan, which is the building block of dense arrays of
overlapping regulation, performs hard-wired combinatorial
decisions governed by the input
functions of the output genes ~\cite{Yuh,Hwa,Setty}.\\

    Here, we address the question of whether a given network motif
appears independently in the network or whether instances of the
motif combine to form larger structures ~\cite{Barabasi_BMC}. If
the latter occurs, what is the function of these larger
structures? Do different networks that share a certain network
motif also share the same structural combinations of that motif?
These questions require analysis of large subgraphs, a
computationally difficult
problem~\cite{Nesetril,Harary,Itzkovitz,Kashtan}. Recently,
efficient algorithms for counting subgraphs based on sampling have
been introduced~\cite{Kashtan}. These algorithms can at present be
effectively used to detect motifs of up to 7-8 nodes. To go beyond
this requires an approach to efficiently define and detect large
structures whose architecture is based on a given motif.

\begin{figure*}
\begin{center}
\includegraphics[width = 130 mm]{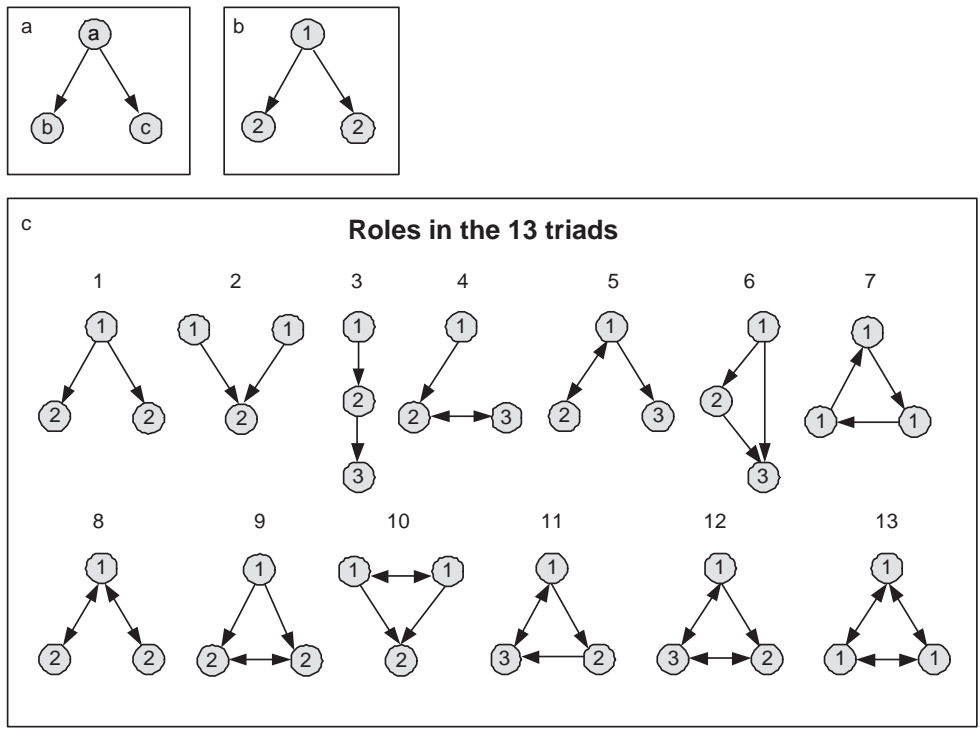}
\end{center}
\caption{\textbf{a.} A directed 3-node subgraph (triad)
\textbf{b.} This triad has two roles. \textbf{c.} Roles in all the
13 types of connected triads. In each triad there are between one
and three roles.} \label{fig1}
\end{figure*}

To address these issues, we present an approach for uniting
related groups of motifs of different sizes into families termed
\emph{motif generalizations}. This allows generalizing from small
motifs to the larger complexes in which they appear. We present an
efficient algorithm to detect motif generalizations. We find that
networks that share the same motif can have different
generalizations of that motif. For example, we find different
generalizations of the FFL motif in transcription, neuronal and
electronic networks. Using mathematical models we analyze the
information-processing functions of the FFL generalization that is
selected in each of these networks.

\section{RESULTS}

\subsection{Node Roles in a subgraph}
We begin by defining \emph{roles} of nodes in a subgraph. A group
of nodes in a subgraph share the same role if there is a
permutation of these nodes, together with their corresponding
edges, that preserves the subgraph structure (See APPENDIX A for
formal definitions). For example, in the v-shaped subgraph in Fig.
1a, nodes b and c can be permuted leaving the structure intact,
whereas nodes a and b cannot. Thus, this subgraph has two roles,
role 1 and role 2 (Fig. 1b). The FFL has three roles (Fig. 1c,
triad 6), whereas the 3-loop (Fig 1c, triad 7) has only one role
(because a cyclic permutation of the three nodes preserves its
structure). The thirteen possible connected-directed triads have
between one and three roles each (Fig. 1c), with a total of 30
different roles.

\begin{figure*}[!hbp]
\begin{center}
\includegraphics[width = 140 mm]{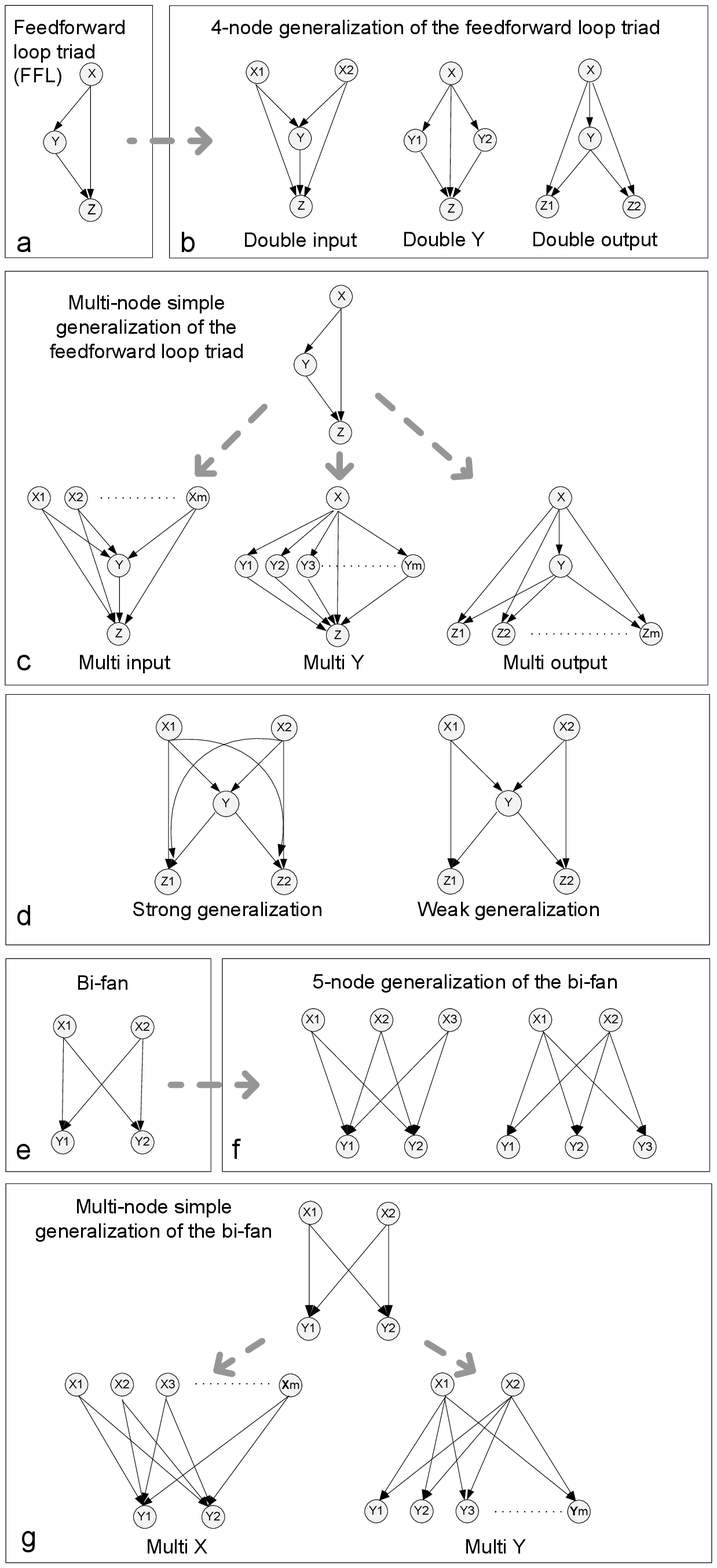}

\end{center}
\caption{\textbf{a.} The feedforward loop triad has three roles: X
(input node), Y (internal - secondary input) node and Z (output
node) \textbf{b.} 4-node simple generalizations of the feedforward
loop. The X node is duplicated to form the double-X
generalization. The Y and Z nodes are duplicated to form the
double-Y and double-Z generalizations respectively. \textbf{c.}
Simple multi-node generalizations of the FFL. \textbf{d.} Strong
and weak generalization rules. A 5-node generalization of the FFL
with two X nodes, one Y node, and two Z nodes. In the strong
generalization every combination of a X,Y,Z triplet of nodes forms
a FFL. \textbf{e.} The bi-fan, a 4-node motif with two roles X
(input role) and Y (output role). \textbf{f.} 5-node simple
generalizations of the bi-fan. In each of the two generalizations
one of the two roles is duplicated. \textbf{g.} Simple multi-node
generalization of the bi-fan: an X or Y node is replicated to form
the multi-input or multi-output bi-fan generalization
respectively.}  \label{fig2}
\end{figure*}

\begin{figure*}
\begin{center}
\includegraphics[height = 50 mm ]{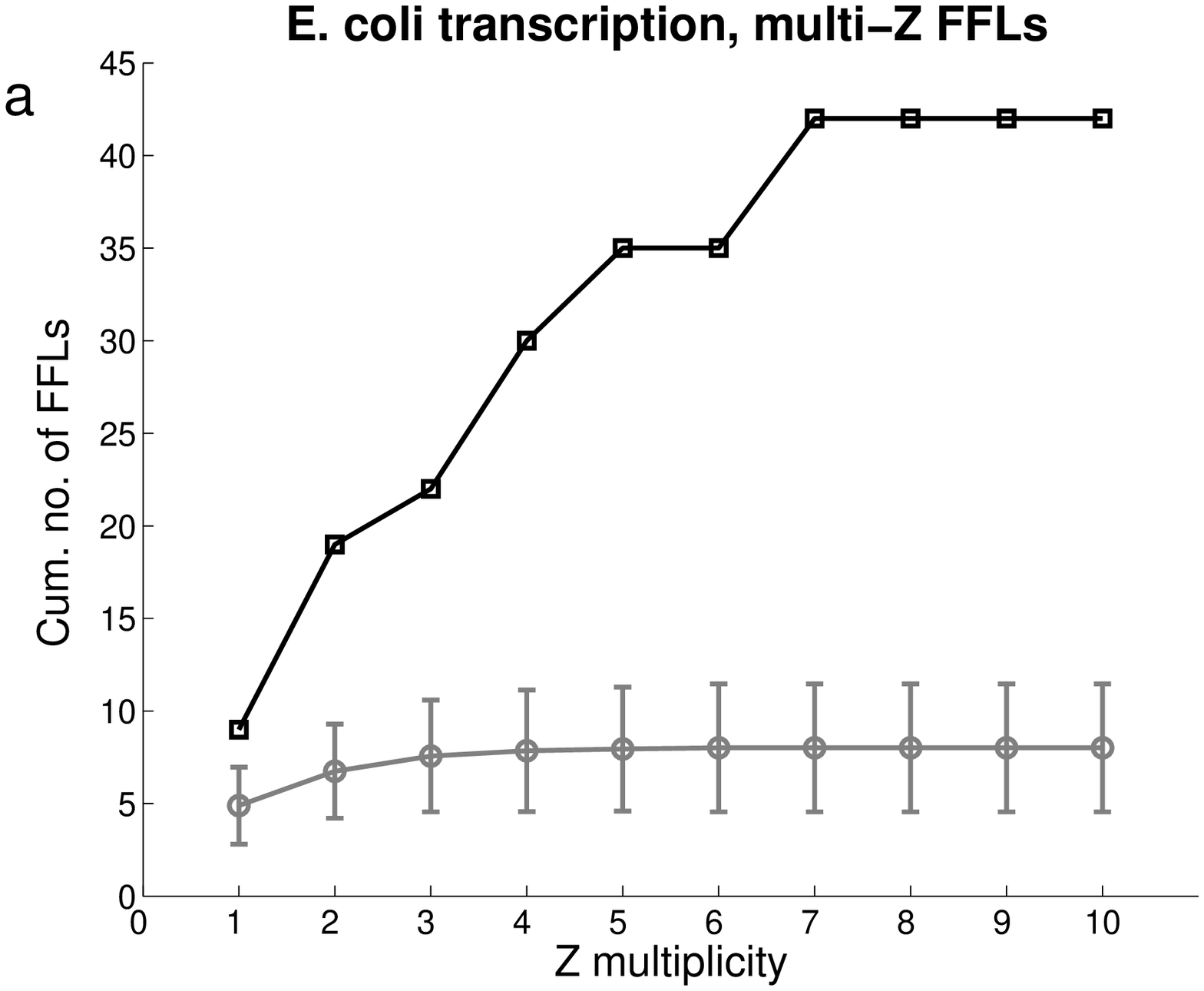}
\includegraphics[height = 50 mm ]{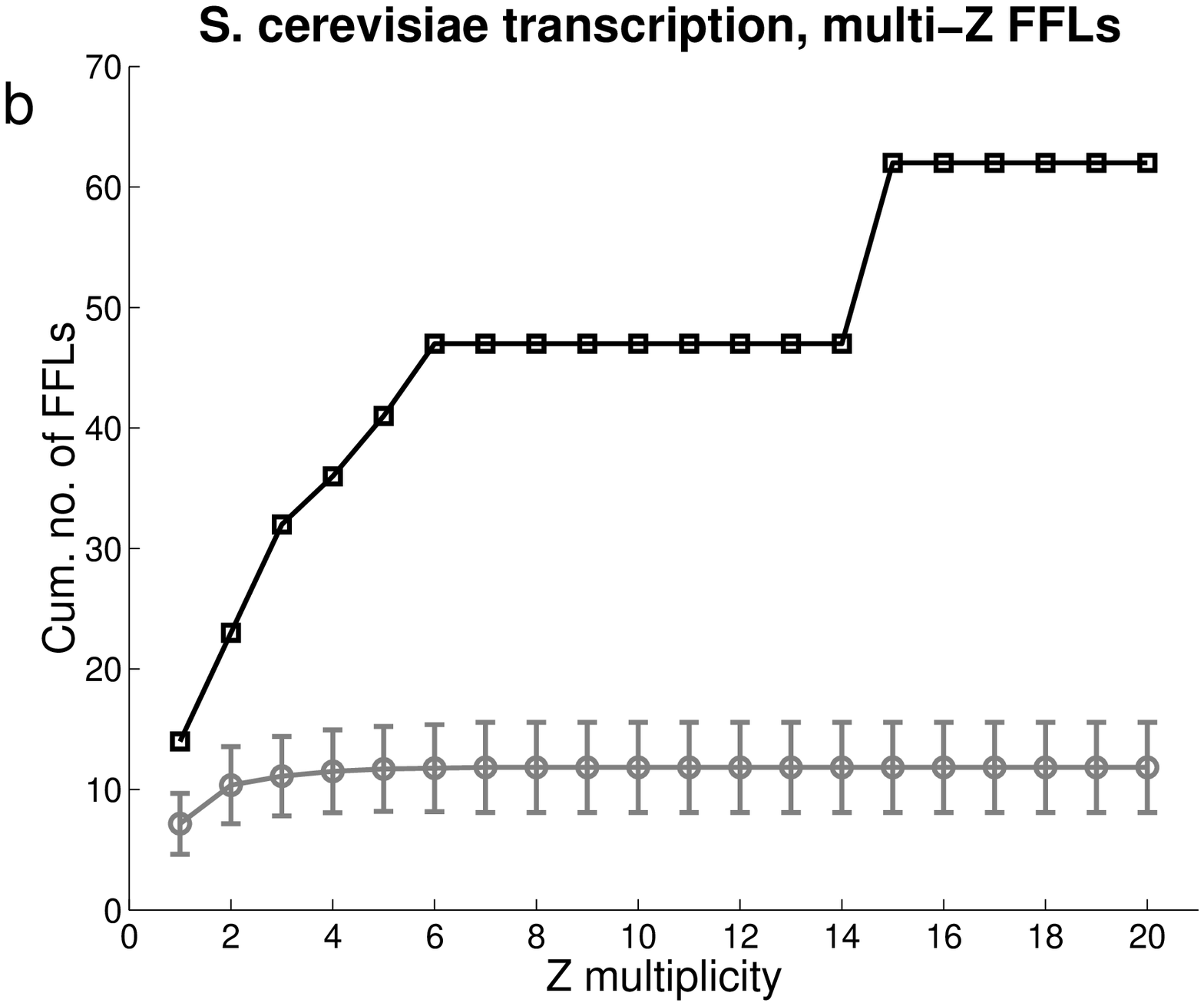}
\includegraphics[height = 50 mm ]{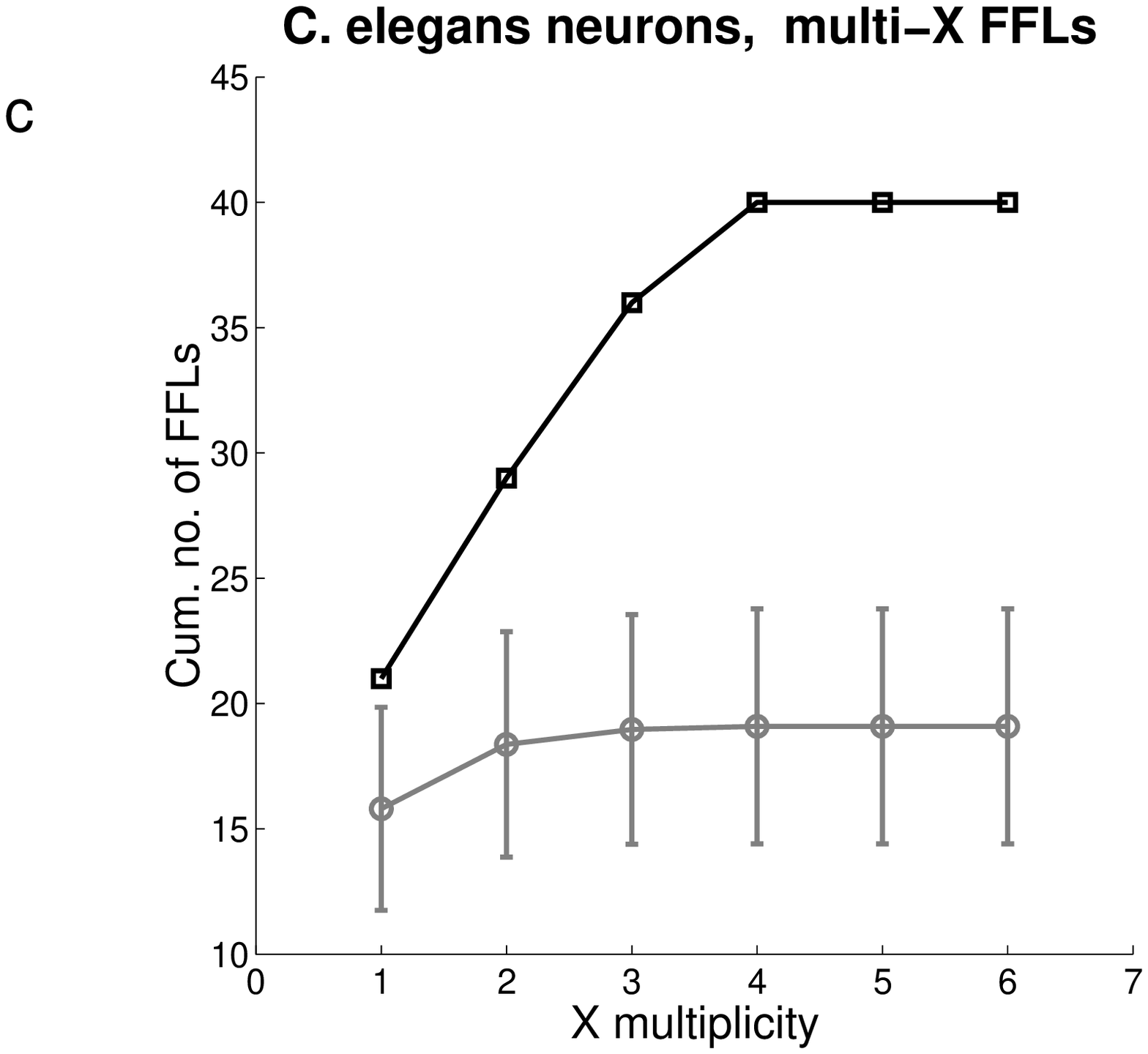}
\end{center}
\caption{\textbf{Statistical significance of motif
generalizations.} The cumulative number of multi-Z FFLs in the
real network (black) and randomized networks - mean $\pm$ SD
(grey) in \textbf{a.} \emph{E. coli} transcription network.
\textbf{b.} \emph{S. cerevisiae} transcription network.
\textbf{c.} The cumulative number of multi-X FFLs in the real and
randomized networks (mean $\pm$ SD) in the \emph{C. elegans}
neuronal network.} \label{fig3}
\end{figure*}

\begin{table*}
\begin{tabular}{l|c|l|l|l|l|l|}
Generalization & Subgraph size &
\multicolumn{2}{c|}{Transcriptional Networks} & Neurons &
Electronic chips \\
\cline{3-6}
&  &  E. coli & yeast & C. elegans & S15850 \\
\hline
basic bi-fan & 4 (2X,2Y) & + (N=209) & + (N=1812) & + (N=126) & + (N=1040)\\
\hline
multi output & 5 (2X,3Y) & + (N=264) & + (N=14857) & + (N=152) & + (N=1990)\\
& 6 (2X,4Y) & + (C=0.015) & + (C=3.5) & +(C=0.17) & + (C=0.28)\\
\hline
multi input & 5 (3X,2Y) & + (N=20) & + (N=81) & + (N=25) & + (N=226)\\
& 6 (4X,2Y) & - (N=0) & + (N=14) & +(C=0.015) & + (C=0.001)\\
\hline
equal multi input-outputs & 6 (3X,3Y) & + (N=6) & + (N=21) & - (N=0) & + (N=301)\\

\end{tabular}
\caption{\textbf{Bi-fan generalizations in different networks.}
(aX,bY) represents the multiplicity of each of the roles in the
generalization (Fig. 2g). '+': Statistically significant
generalizations, '-': non-significant generalizations. Number of
appearances (N), or concentration (x10$^{-3}$) (C)
 ~\cite{Kashtan} are listed.} \label{Table1}
\end{table*}

\subsection{Subgraph Topological Generalizations}
We now define subgraph topological generalizations based on node
roles. Subgraph topological generalizations are extensions of a
subgraph to a family of larger subgraphs which share its basic
structure. Consider the FFL (Fig. 2a). For this 3-node subgraph we
define three simple generalizations to the level of 4 nodes (Fig.
2b). In each simple generalization a single role and its
connections are duplicated. In the first simple generalization,
the X role and its connections are duplicated. This generalization
is termed double-X FFL or double-input FFL. The other two
generalizations are obtained by duplicating the Y or Z roles. This
replication process can be continued, leading to higher-order
motif generalizations, the multi-X (multi-input), multi-Y and
multi-Z (multi-output) FFL generalizations (Fig. 2c).\\
    More complex generalizations can be obtained by replicating more
than one of the roles. For example, duplicating both the X and Z
roles yields five-node generalizations (Fig 2d). When replicating
more than one role (and in some cases replicating even a single
role), one can define two kinds of generalizations: in strong
generalizations, every X,Y,Z triplet forms a FFL. In weak
generalizations, every node participates in at least one FFL, but
not all possible FFLs are formed (Fig. 2d).\\
    This procedure of generalization can be applied to any subgraph
(see formal definition in APPENDIX B). For example simple
generalizations of the 4-node bi-fan are shown in Fig. 2e-g. We
now describe the statistical-significance of the generalizations
of the motifs found in various networks.

\subsection{Network Motifs Topological Generalizations}
While enumerating all subgraphs of a given size is a difficult
task, enumerating generalizations of a given subgraph can be
performed efficiently by an algorithm described in APPENDIX C. The
algorithm is based on using the appearances of the basic subgraph
as nucleation points for a search for its generalizations. As an
example, we applied this algorithm to networks in which the FFL
and bi-fan are motifs, to ask whether any of the possible FFL or
bi-fan generalizations occur significantly in the networks
(APPENDIX C). In the transcription networks of \emph{E. coli}
~\cite{Shen-Orr} and \emph{S. cerevisiae} ~\cite{Milo} we find
that the multi-Z FFL generalization is highly significant (Fig.
3a,b). The other two possible simple generalizations are not
significant (in the \emph{E. coli} network, multi-X's and
multi-Y's do not occur at all, in the \emph{S. cerevisiae} network
both appear only twice). An example of a multi-Z FFL in the
\emph{E. coli} transcription network, the maltose utilization
system, is shown in Fig. 4a. In each multi-Z FFL, the different
genes (Z roles) share a common biological function (as shown in
tables 2 and 3 that list all multi-Z FFL
complexes in the  \emph{E. coli} and \emph{S. cerevisiae} networks).\\

\begin{figure}[!hbp]
\begin{center}
\includegraphics[width = 86 mm, height = 160 mm ]{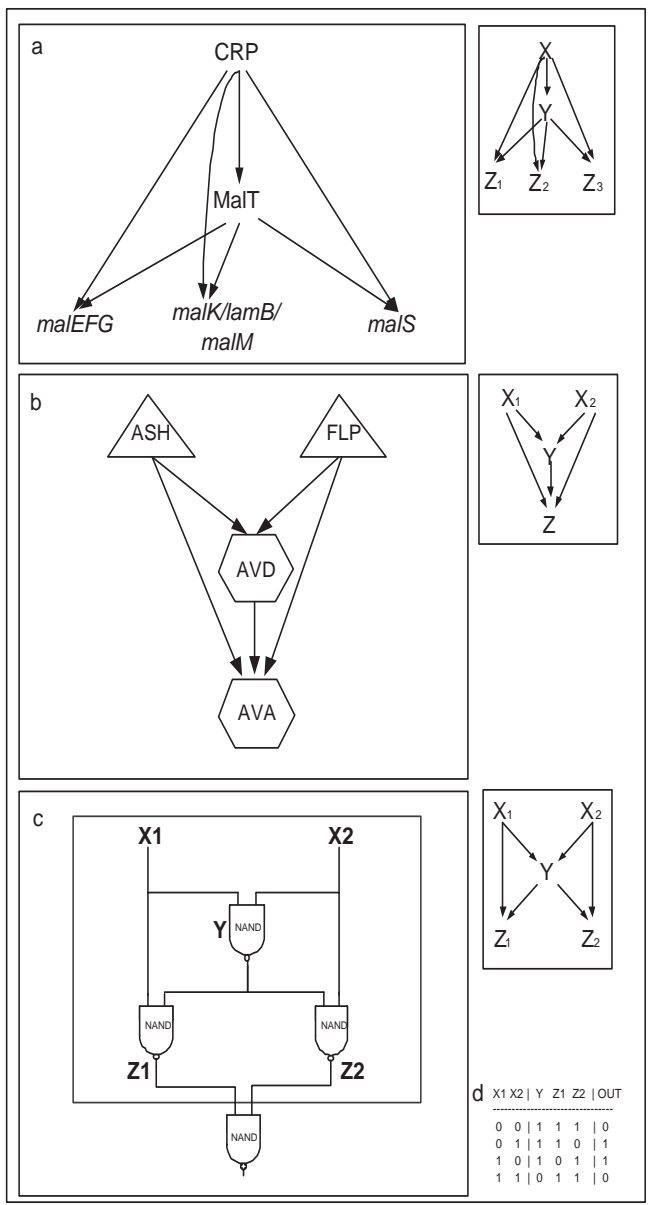}
\end{center}
\caption{\textbf{The FFL generalizations found in biological and
technological networks.} \textbf{a.} An example of a three-Z FFL
in the transcription network of \emph{E. coli}, maltose
utilization system. The activator CRP senses glucose starvation,
MalT senses maltotriose, and malEFG, malK and malS participate in
maltose metabolism and transport. \textbf{b.} An example of a
double-X FFL in the locomotion neural circuit of \emph{C.
elegans}. AVA and AVD are ventral cord command interneurons. AVD
functions as modulator for backward locomotion. AVA functions as
driver cell for backward locomotion. ASH and FLP are head sensory
neurons sensitive to noxious chemicals and nose touch. \textbf{c.}
A generalized form of the FFL (2X,Y,2Z) found in forward-logic
electronic chips. This 5-node structure appears as a part of a
6-node module, which implements XOR (Exclusive OR) using 4 NAND
gates. \textbf{d.} Truth table of the circuit described in c (a
(2X,Y,2Z) FFL generalization with additional NAND gate at the
output). There are 2 input bits X1 and X2 and a single output bit
which is equal to (X1 \textrm{XOR} X2).} \label{fig4}
\end{figure}

In the network of synaptic connections between neurons in \emph{C.
elegans} ~\cite{White,AY,Milo}, we find a different FFL
generalization: the multi-X FFL (Fig. 3c). This structure occurs
29 times in the network, with upto 4 inputs. Multi-Y and multi-Z
FFLs are found in far smaller numbers (double-X's and double-Y's
FFL appear 3 times each) ~\cite{Neurons_Comment}. An example of a
multi-X FFL in the locomotion control circuit of \emph{C. elegans}
is shown in Fig. 4b.\\

In networks of connections between logic gates in forward-logic
electronic chips ~\cite{Brglez,Milo,Sole} we find no simple
generalization of the FFL. These electronic circuits do, however,
show a complex FFL generalization - a structure with two Xs, a
single Y and two Zs (a weak generalization, Fig. 4c). In the five
forward-logic electronic chips we have analyzed, 70 percent to 100
percent of
the FFLs are embedded in instances of this 5-node structure.\\

The most prominent 4-node network motif in these networks is the
bi-fan ~\cite{Milo} (Fig. 2e). The bi-fan has two roles and
therefore two simple generalizations (Fig. 2g). We find that both
simple generalizations of the bi-fan (multi-output and
multi-input) are significant in the transcription, neuronal and
electronic networks (Table 1). The multi-output bi-fan
generalizations are more significant and the maximal Y
multiplicity is higher than the maximal X multiplicity in all
these networks. In these networks we find structures of
multi-output bi-fan with 10 Ys and more, while multi-input bi-fan
do not exceed 6 input X nodes.

\subsection{Functions of multi-output FFL generalization in transcription networks}
The function of the FFL depends on the signs of the interactions
(positive or negative regulation), on their strengths and on the
functions that integrate multiple inputs into each node. In the
case of positive regulation, the 3-node FFL has been shown to
function as a persistence detector ~\cite{Shen-Orr}: it filters
out short input stimuli to X, and responds only to persistent
signals. On the other hand, it responds quickly to OFF steps in
the input to X ~\cite{Shen-Orr,Mangan}. With other sign
combinations, the 3-node FFL can function as a pulse-generator or
response accelerator ~\cite{Mangan,Weiss}. These functions apply
to a wide range of interaction strengths,
and to both AND and OR-like input functions.\\

\begin{figure}[!hbp]
\begin{center}
\includegraphics[width = 55 mm, height = 65 mm ]{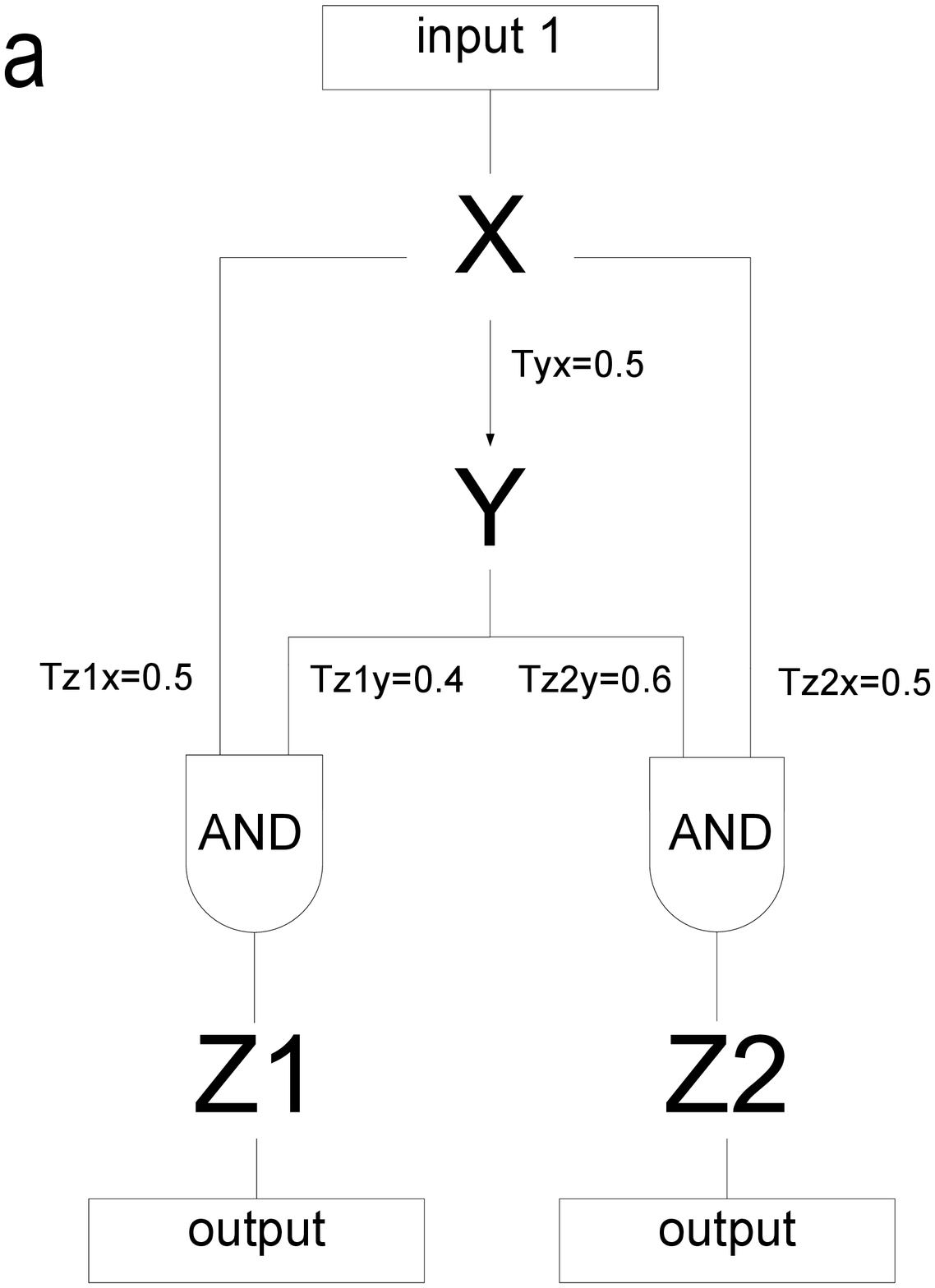}
\includegraphics[width = 80 mm, height = 60 mm ]{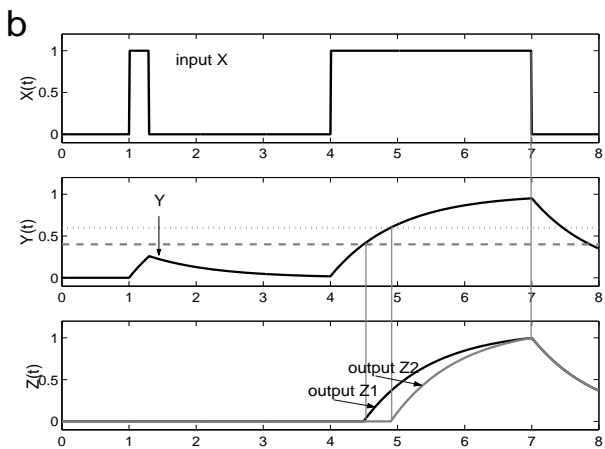}
\end{center}
\caption{Kinetics of a double-output FFL generalizations following
pulses of stimuli. \textbf{a.} A double-output FFL with positive
regulation and AND-logic input function for $Z_{1}$and $Z_{2}$.
Numbers on the arrows are activation thresholds. \textbf{b.}
Simulated kinetics of the double-output FFL in response to a short
pulse and a long pulse of X activity. The dashed and dotted
horizontal lines represent the activation thresholds $T_{z_{1}y}$
and $T_{z_{2}y}$. $\alpha=1$ was used.} \label{fig5}
\end{figure}

Here, we studied the functions of the generalizations of the FFL.
We begin with the multi-output FFL, which is the generalization
that is significant in transcription networks. The multi-output
FFL has a single input node X, a single internal node Y (secondary
input) and a number of output nodes $Z_{1}..Z_{m}$ (Fig. 2c,4a).
The arrows in the FFL diagram should be assigned numbers
representing the strength of the interaction of the transcription
factors (TFs) X and Y with the promoters of the various Z-genes
~\cite{Ronen}. These numbers correspond to the activation or
repression coefficients of each gene (the concentration of the TF
required for 50 percent effect
~\cite{Savageau,Ronen,McAdams1998}). Here, we consider for
simplicity the most common case, that of FFLs with positive
regulation ~\cite{Mangan}. We employ a simple model of the
dynamics of this circuit ~\cite{Shen-Orr}. $X(t)$ is the activity
of the transcription factor $X$, $Y(t)$ of $Y$, $Z_{j}(t)$ is the
concentration of the gene product $Z_{j}$.
The dynamics of  transcription factor Y and the output gene products $Z_{j}$ is given by\\\\

$dY/dt=F(X,T_{yx})- \alpha Y$\\\\

$dZ_{j}/dt=F(X,T_{z_{j}x})F(Y,T_{z_{j}y})- \alpha Z_{j}$\\\\

Where $\alpha$ is the protein lifetime
~\cite{Rosenfeld,Rosenfeld_2003} and $T_{yx}$, $T_{z_{1}x}$,
$T_{z_{2}x}$, $T_{z_{1}y}$, $T_{z_{2}y}$ are the activation
thresholds of the various genes (Fig. 5a). For simplicity we use a
sharp activation function, $F(U,T)=1$ if $U>T$ and 0 otherwise.
The qualitative results apply also to Michaelis-type activation
functions. These equations can be solved analytically, yielding
piecewise exponential dynamics in response to step like activation
profiles of X. We find that the multi-output FFL can encode a
temporal order of expression of the Z genes, by means of different
activation thresholds $T_{z_{j}y}$ for each of the output genes
(Fig. 5a,b). This temporal ordering feature is shared with another
common network motif, the single-input module
~\cite{Shen-Orr,Ronen,Zaslaver}. Indeed, high resolution
expression measurements on the flagella multi-output FFL (in
\emph{E. coli}) showed that the class 2 flagella genes, which are
regulated by a feedforward loop, are activated in a temporal order
that corresponds to the functional order of the gene product in
the assembly of the flagellar motor ~\cite{Kalir,Kalir_Cell}.\\

The timing of activation of gene j following a step activation of X is\\

$\tau_{j}=-\alpha^{-1} \ln (1-T_{z_{j}y}/Y_{max}) $\\\\

The rise time of the different genes can be tuned by
$T_{z_{j}y}/Y_{max}$, where $Y_{max}$ is the maximal concentration
of $Y$. Note that $T_{z_{j}y}$ can be easily tuned during
evolution , for example by mutations in the binding site of $Y$ in
the $Z_{j}$ promotor ~\cite{Setty,Kalir_Cell}. The Z gene with the
lowest activation threshold is turned on first after the
stimulation of X. Furthermore, the multi-Z FFL can act as a
persistence detector for all of the output genes (Fig. 5b): the Z
genes are expressed only if the input stimulus to X is present for
a long enough time. The minimal time that a saturating X stimulus
needs to be present to activate gene j is equal to $\tau_{j}$.
Thus this FFL generalization preserves the functionality of the
original FFL motif.  The turn-off order of the Z genes upon a
gradual decay of X activity can be separately controlled by the
activation coefficients of the X TF, $T_{z_{j}x}$
~\cite{Kalir_Cell}. Thus different turn on and turn off orders of
the $Z_{j}$ genes can in principle be achieved. In summary, the
multi-output FFL preserves the functionality of the simple FFL,
and in addition can encode
temporal expression programs among the different Z genes.\\

\begin{figure}[!hbp]
\begin{center}
\includegraphics[height = 75 mm ]{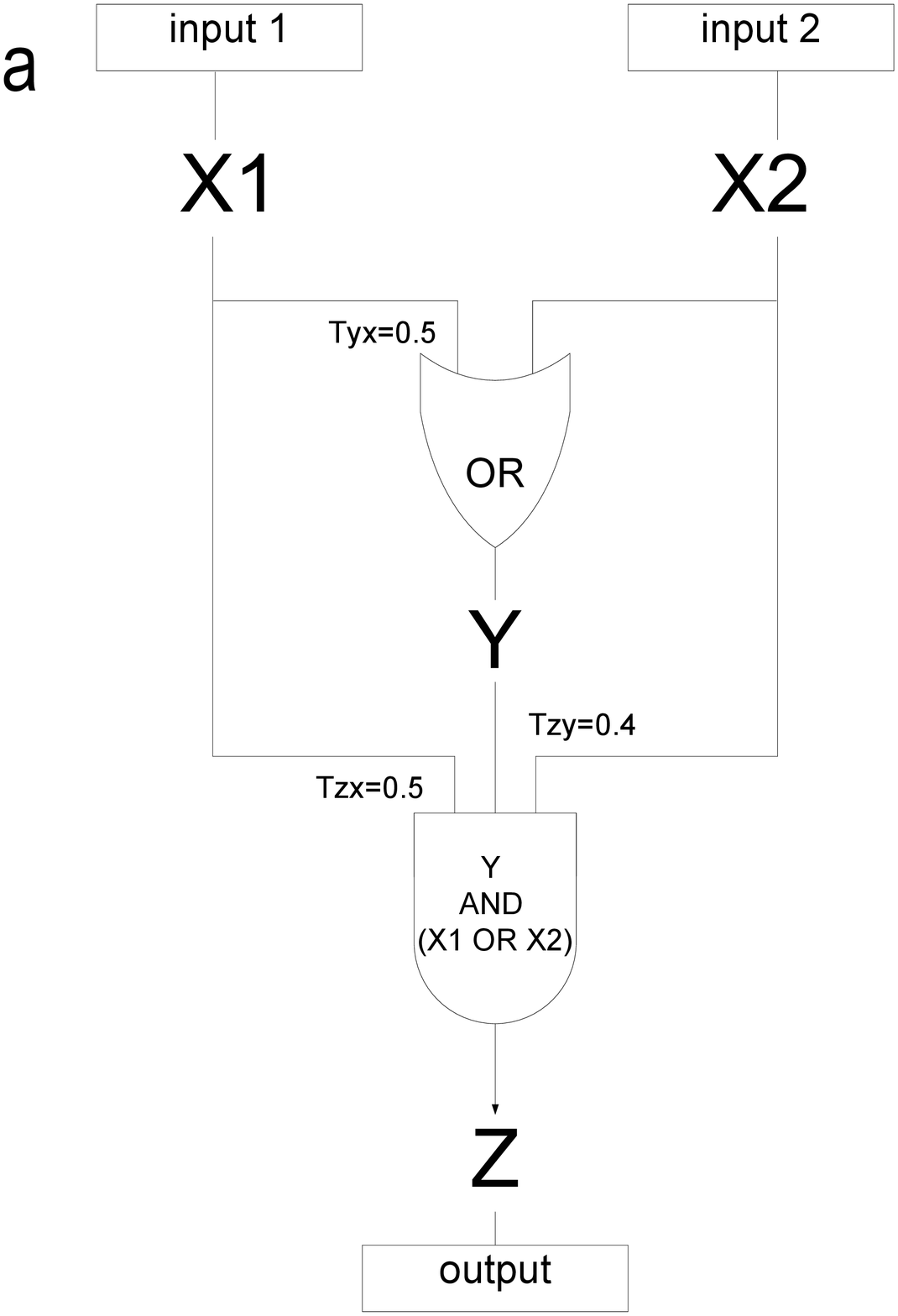}
\includegraphics[width = 80 mm, height = 60 mm ]{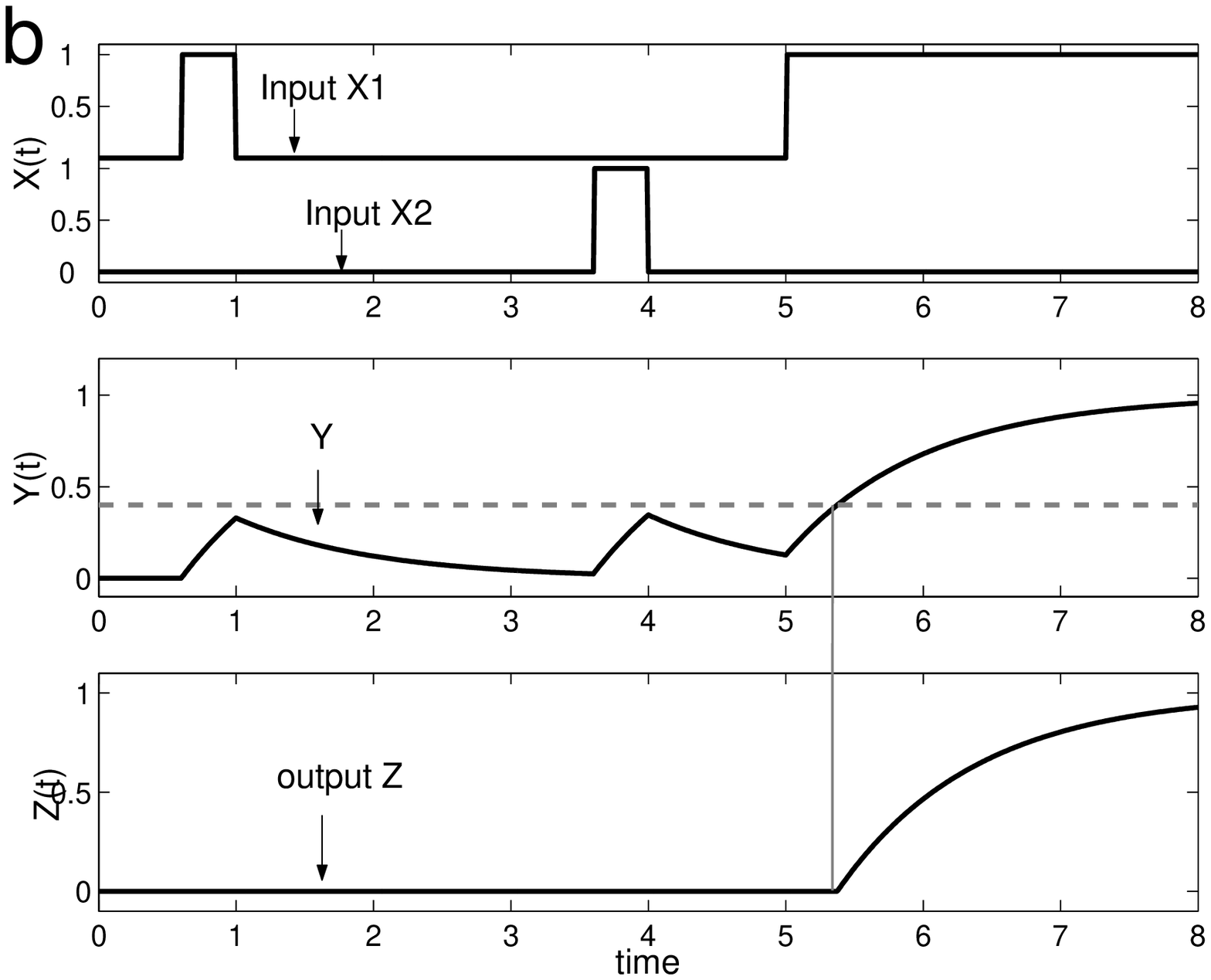}
\includegraphics[width = 80 mm, height = 60 mm ]{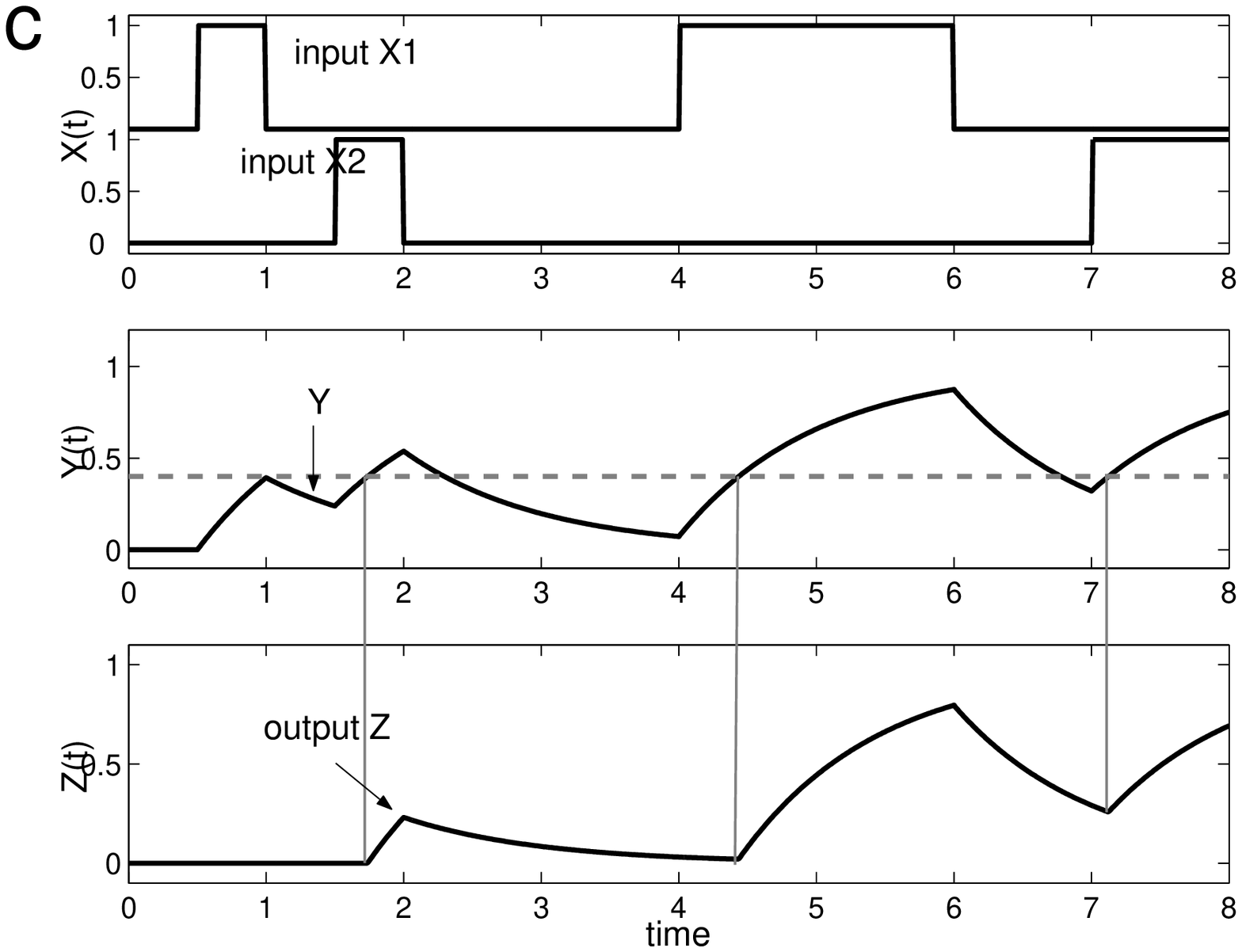}

\end{center}
\caption{Kinetics of a double-input FFL generalization following
pulses of stimuli. \textbf{a.} A double-input FFL. Input functions
for Y and Z, and the activation thresholds, are shown as gates and
numbers on the arrows.  \textbf{b.} Simulated kinetics of the
two-input FFL, with short well-separated stimuli pulses of $X_{1}$
and $X_{2}$, followed by a persistent $X_{1}$ stimulus.
\textbf{c.} Simulated kinetics of the double-input FFL, with short
$X_{1}$ stimulus followed rapidly by a short $X_{2}$ stimulus
pulse. The dashed horizontal line corresponds to the activation
thresholds for $Y$, $T_{zy}$. $\alpha=1$ was used.} \label{fig6}
\end{figure}

\subsection{Functions of multi-input FFL generalization in neuronal networks}
A different FFL generalization, multi-input FFL, is found in the
neuronal network of C. elegans. In general, the function of this
circuit depends on the signs on the arrows and on two
input-functions (gates): one input function integrates the
multiple X inputs to Y, and the
other integrates the inputs from Y and $X_{1}..X_{m}$ to Z. (Fig. 6a) \\

We analyzed the dynamics of  one possible two-input FFL, where the
input-function governing the Y node is an OR gate, $X_{1}$
\textrm{OR} $X_{2}$, and the input-function of the Z node is $Y$
\textrm{AND} $(X_{1}$ \textrm{OR} $X_{2})$ (Fig. 6a,b,c). This
choice of input-functions ensure that both $Y$ and either $X_{1}$
or $X_{2}$ are needed for Z to be activated to a level that allows
activation of its downstream (post synaptic) neurons or muscle
cells (as is the case, for example, in the circuit of Fig. 4b, in
which ablation of the neuron AVD results in loss of sensory input
to the neuron AVA ~\cite{Chalfie}). These input functions could in
principle be implemented by simple neurons which integrate
weighted inputs. The input function of $Z$, for example,
represents strong synapses from $Y$ and weaker ones from $X_{1}$
and $X_{2}$. It is important to note that the simplest equations
that describe transcription networks also describe neurons with
graded potential and no spiking (as \emph{C. elegans} neurons are
thought to be ~\cite{Hope,Goodman} ). In the case of neurons,
$X_{i}(t)$, $Y(t)$ and $Z(t)$ represent neuron membrane
potentials. The activation dynamics of the circuit in Fig. 6a are\\\\

$dY/dt=F(X_{1}+X_{2},T_{yx})- \alpha Y$\\\\

$dZ/dt=F(Y,T_{zy})[F(X_{1}+X_{2},T_{zx})]- \alpha Z$\\\\

Here $\alpha$ is the relaxation rate of the neurons' membrane
potential, and the synaptic activation thresholds are
$T_{yx}$,$T_{zx}$, $T_{zy}$. \\

This model shows that the circuit can act as a persistence
detector for both $X_{1}$ and $X_{2}$ (Fig. 6b). In the locomotion
neuronal circuit example (Fig 4b), the FFL circuit could elicit
backward motion only if the stimulation of one of the sensory
neurons is longer than a threshold duration $\tau$ determined
by the parameters of the circuit.\\\\

$\tau=-\alpha^{-1} \ln (1-T_{zy}/Y_{max}) $\\\\

A transient stimulation would not be enough to elicit backward
motion. Furthermore, we find that sufficiently closely spaced
short pulses of $X_{1}$ and $X_{2}$ can elicit a response,
\emph{even if each pulse alone can not} (Fig. 6c). This highlights
a 'memory-like' function of Y, which can store information from
recent stimulations over its relaxation time. In the basic 3-node
FFL, Y can store information about recurring pulses of X. In the
multi-input FFL, Y can store information from multiple inputs
(Fig. 6c gives an example), and increase sensitivity to one input
if the other input has recently been detected. Generally, if the
summed input of the input-nodes $X_{j}$ to node Y is
$S(t)=F(x_{1}+x_{2},T_{yx})$, Z is activated when Y activity
exceeds the threshold $T_{zy}$\\\\

$Y(t)=e^{- \alpha t} \int_{0}^{t} S(t^{\prime}) e^{\alpha t^{\prime}} d t^{\prime} > T_{zy}$\\\\

(where $Y(t=0)=0)$, showing that node Y effectively
integrates the inputs over a time scale of $1/\alpha$.\\

\subsection{Function of FFL generalization in electronic chips}
Forward-logic electronic chips are networks in which nodes
represent logic gates. These circuits are optimized to perform a
hard-wired logical function between input and output nodes.
Forward-logic chips, taken from an engineering database (ISCAS89),
were previously found to display the FFL network motif
~\cite{Milo}. Here we find that they display a specific
generalization of the FFL, with two input and two output nodes
(Fig. 4c). Analyzing the appearances of this pattern, we find that
this 5-node generalized FFL motif is part of a commonly used
engineering module built of 4 NAND gates, which implements XOR
(exclusive OR) logic on the two inputs ~\cite{Hansen} (see truth
table in Fig. 4d).\\

\section{DISCUSSION}
 This study presented a systematic approach for defining and
detecting generalizations of network motifs. Motif generalizations
are families of subgraphs of different sizes which share a common
structural theme, and which appear significantly more often in the
network than in randomized networks. The generalizations are
produced by replicating nodes in a basic motif structure. The
generalizations often preserve the functionality of the network
motif on which they are based, because they preserve the roles of
nodes in the motif (for example, by replicating input or output
nodes). We presented an efficient algorithm for detecting motif
generalizations. We find that different networks which display the
same motifs can show very different generalizations of these
motifs. We also demonstrated using simple models that these
generalized motifs can carry out specific information processing
functions. These functions can in principle be tested
experimentally in transcription and
neuronal systems.\\

The two sensory transcription networks, from a prokaryote
(\emph{E. coli}) and a eukaryote (\emph{S. cerevisiae}), showed
the same generalization of the FFL: both networks display the
multi-output FFL generalization ~\cite{Shen-Orr,Barabasi_BMC}. The
other two generalizations, multi-input and multi-Y, are not found
significantly in these transcription networks.  Multi-output FFL
complexes are found throughout the transcription networks in
diverse systems (tables 2,3). The X role is usually a global
transcription factor which controls many genes, the Y role is
usually a 'local' transcription factor which controls specific
systems, and the Z nodes are the regulated genes which share a
specific function. Often, multi-output FFLs in \emph{E. coli} that
respond to specific stimuli have a non-homologous multi-output FFL
counterpart in yeast which responds to similar stimuli. The fact
that the genes in these circuits are not evolutionary related,
whereas their connectivity patterns are the same in the two
organisms, suggests convergent evolution to the same regulation
pattern ~\cite{Milo,Conant}. Examples include systems that respond
to carbon limitation, drugs, and nitrogen starvation in both
organisms (tables 2,3). Multi-output FFLs can also appear in
systems that make up a protein machine, for example, a
multi-output FFL in \emph{E. coli} controls genes whose products
make up the flagellar basal-body motor ~\cite{Kalir} (X=flhDC,
Y=fliA, Z= class 2 flagella genes). We find that the multi-output
FFL can serve as a persistence detector for all the outputs. In
addition it can generate temporal orders of output gene expression ~\cite{Kalir_Cell}.\\

A different FFL generalization, the multi-input FFL, is found in
the neuronal synaptic wiring of \emph{C. elegans}. This network is
found to chiefly display the multi-input FFL (Fig. 2c), and not
the other two generalizations. The multi-input FFL has a number of
input nodes $X_{1}..X_{m}$, a single internal node Y (secondary
input) and a single output node Z. As an example we have mentioned
the backward locomotion control circuit of the worm. This circuit
is governed by two ventral-cord command interneurons AVD and AVA
~\cite{Chalfie,Hope}. These two neurons are linked in a
multi-input FFL with several input neurons, such as ASH and FLP
(Fig. 4b), which are head sensory neurons sensitive to nose touch
and noxious chemicals ~\cite{Chalfie,Hope}. This circuit
implements an avoidance reflex, eliciting backward motion in
response to head stimulation. We find that the multi-input FFL can
serve as a persistence detector for each input. In addition, it
can serve as coincidence detector for weak inputs, firing only if
short stimuli from two or more different inputs
occur within a certain time of each other.\\

A different FFL generalization, with two inputs and two outputs,
appears in a class of electronic circuits. This motif
generalization functions within a XOR gate. This demonstrates that
network motifs and their generalizations can be used to detect
basic functional
building block of a network without prior knowledge.\\

    Motif generalizations cover a substantial portion of the high-order motifs
in various biological and technological networks we have studied.
However, motifs generalizations in the present form do not cover
all possible types of families of structures that share similar
architectural themes. It would be important to find additional
rules for defining families of motifs beyond the current notion of
motif generalization by role replication. Motifs and their
generalizations can help us understand the design principles of
complex networks by defining functional building blocks whose
function can be tested experimentally.\\

    To summarize, this study presented topological generalizations of
network motifs, and an efficient algorithm to detect them. We
found motif generalizations in several real-world networks.
Networks that share the same motif were found to exhibit different
generalizations of that motif. We demonstrated theoretically that
the generalized motifs in biological networks can carry out
information-processing functions.

\acknowledgments We thank all members of our lab for discussions.
We thank NIH, Israel Science Foundation and Minerva for support.
N.K. was supported by Ernst and Anni Deutsch-Promotor Stiftung
foundation for an MSc fellowship. R.M. was supported by Horowitz
complexity science foundation PhD fellowship.


\appendix

\section{Roles in a subgraph - formal definition}

We classify nodes in a subgraph into structurally equivalent
classes. Each class represents a role. The measure of structural
equivalence that we use here is automorphic equivalence
~\cite{Wasserman,Lorrain,Winship,Winship&Mandel,Everett}. Let
$S=(V_{s},E_{s})$ be a subgraph, then an automorphism is a
one-to-one mapping, $\tau$ ,from $V_{s}$ to $V_{s}$ , such that
$(v_{i},v_{j}) \in E_{s}$ if and only if $(\tau(v_{i}),\tau(v{j}))
\in E_{s}$. Two nodes $v_{i}$ and $v_{j}$ are automorphically
equivalent if and only if there is some automorphism, $\tau$ ,
that maps one of the nodes to the other $(\tau(v_{i})=v_{j})$. For
each subgraph S, we classify all its n nodes into roles by
examining structural equivalence of all possible pairs of the
nodes. By the transitivity of automorphic equivalence, we are
guaranteed to get a partition of the nodes into distinct roles.
This concept can be readily generalized for networks with weights
on the edges or with different types of nodes.

\section{Subgraph generalization - formal definition}
Let S be the basic subgraph where $r_{1}..r_{L}$ are the set of
roles of S with multiplicity $(d_{1},..,d_{L})$ respectively.
simple generalization of S is a subgraph which is formed by
replication of a single role $r_{i}$ and its edges to preserve the
roles connectivity of S. Note that in a simple generalization only
a single role is replicated. A generalized form of a subgraph is
defined by a pair$(M,V^{L})$ where M is an $L \times L$ image
matrix, which describes the connectivity between roles. $M[i,j]=1$
if there is an edge between role $i$ and $j$ ($i$ is not equal to
$j$), and $M[i,j]=0$ otherwise. $M[i,i]=0$ if there is no edge
between every two nodes of role i, $M[i,i]=1$ if there is a single
edge, and $M[i,i]=2$ if there is a mutual edge. $V^{L} \in N^{L}$
is an L-dimensional vector which defines the multiplicity of each
role. The FFL which is an example of a basic subgraph, is
represented by $(M_{FFL},(1,1,1))$ where
\begin{displaymath}
\mathbf{\scriptsize{M_{FFL}}} = \left( \begin{array}{c|c|c}
$\scriptsize{0}$ & $\scriptsize{1}$ & $\scriptsize{1}$ \\
\hline
$\scriptsize{0}$ & $\scriptsize{0}$ & $\scriptsize{1}$ \\
\hline $\scriptsize{0}$ & $\scriptsize{0}$ & $\scriptsize{0}$
\end{array}\right)
\end{displaymath}
and the vector $(1,1,1)$ describes the roles multiplicity: in the
basic FFL each of the three roles X,Y,Z appears once. A FFL with
two output nodes is represented by the pair $(M_{FFL},(1,1,2))$. A
FFL with m output nodes (m Z-role nodes) is represented by
$(M_{FFL},(1,1,m))$ (Fig 2c). Such a generalization has only one
degree of freedom - the multiplicity of the Z role in the
structure. There are cases, such as multiplicity of more than one
role, where we need additional definition in order to distinguish
between different types of structures. For this we define the
generalization rule r. We define two possible generalization
rules: a strong generalization rule and a weak generalization
rule. An example of a strong and weak $(M_{FFL},(2,1,2))$
generalization is illustrated in Fig 2d. If S is the basic n-node
subgraph with set of L roles represented by the multiplicity
vector $(d_{1},..,d_{L})$ then a \emph{basic n-node set} is every
set of n nodes in the structure that consists of $d_{i}$ nodes of
role i (for all $1\leq i \leq L$). For example every set of three
nodes in the multi output FFL, consisting of the X node, Y node
and one of the Z role nodes, is a \emph{basic n-node set}. A
strong generalization rule, $r_{s}$, requires that every
\emph{basic n-node set} in the structure forms the basic subgraph
S. A weak generalization rule, $r_{w}$, requires that every node
in the structure participates in at least one \emph{basic n-node
set} (Fig. 2d). Note that weak generalization can represent more
than one unique structure of a given size.

\section{Algorithm for motif generalizations detection}

We begin by finding the network motifs (significant subgraphs) of
size n (usually n=3-4) in the network as described in
~\cite{Shen-Orr,Milo}(application and source code are available at
http://www.weizmann.ac.il/mcb/UriAlon/). For each motif, for each
of its roles, we prepare a list of all the nodes that play that
role. We perform a search for all of the generalizations of each
motif using its appearances in the network as starting point. This
search reduces computation time and enables the detection of
significant generalization forms of the basic motifs, which are
beyond reach using algorithms
that attempt to enumerate all subgraphs of a given size.\\

\begin{table*}
\begin{tabular}{|c|l|l|l|l|l|}
\hline
\tiny{Complex} & \scriptsize{Id}. & \scriptsize{X} & \scriptsize{Y} & \scriptsize{Z} & \scriptsize{Function}\\
\scriptsize{size} &  &  &  &  & \\
\hline
& \scriptsize{1} & \scriptsize{arcA} & \scriptsize{appY} & \scriptsize{appCBA} & \scriptsize{Anaerobic/stationary phase}\\
& \scriptsize{2} & \scriptsize{crp} & \scriptsize{fucPIKUR} & \scriptsize{fucAO} & \scriptsize{Fucose utilization}\\
& \scriptsize{3} & \scriptsize{crp} & \scriptsize{fur} & \scriptsize{cirA} & \scriptsize{Iron citrate uptake}\\
& \scriptsize{4} & \scriptsize{crp} & \scriptsize{galS} & \scriptsize{mglBAC} & \scriptsize{Carbon utilization}\\
\scriptsize{1} & \scriptsize{5} & \scriptsize{crp} & \scriptsize{malI} & \scriptsize{malXY} & \scriptsize{Maltose utilization}\\
& \scriptsize{6} & \scriptsize{crp} & \scriptsize{melR} & \scriptsize{melAB} & \scriptsize{Melibiose utilization}\\
& \scriptsize{7} & \scriptsize{hns} & \scriptsize{flhDC} & \scriptsize{fliAZY} & \scriptsize{Flagella regulation}\\
& \scriptsize{8} & \scriptsize{metJ} & \scriptsize{metR} & \scriptsize{metA} & \scriptsize{Methionine biosynthesis}\\
& \scriptsize{9} & \scriptsize{ompR-envZ} & \scriptsize{csgDEFG} & \scriptsize{csgBA} & \scriptsize{Osmotic stress response}\\
\hline
& \scriptsize{10} & \scriptsize{crp} & \scriptsize{caiF} & \scriptsize{caiTABCDE} & \scriptsize{Carnitine metabolism}\\
&  &  &  & \scriptsize{fixABCX} & \\
& \scriptsize{11} & \scriptsize{crp} & \scriptsize{nagBACD} & \scriptsize{manXYZ} & \scriptsize{Carbon utilization}\\
&  &  &  & \scriptsize{nagE} & \\
\scriptsize{2} & \scriptsize{12} & \scriptsize{himA} & \scriptsize{ompR-envZ} & \scriptsize{ompC} & \scriptsize{Osmotic stress response}\\
&  &  &  & \scriptsize{ompF} & \\
& \scriptsize{13} & \scriptsize{rpoN} & \scriptsize{fhlA} & \scriptsize{fdhF} & \scriptsize{Formate hydrogen lyase system}\\
&  &  &  & \scriptsize{hycABCDEFGH} & \\
& \scriptsize{14} & \scriptsize{rpoN} & \scriptsize{glnALG} & \scriptsize{glnHPQ} & \scriptsize{Nitrogen utilization}\\
&  &  &  & \scriptsize{nac} & \\
\hline
\scriptsize{3} & \scriptsize{15} & \scriptsize{crp} & \scriptsize{malT} & \scriptsize{malEFG} & \scriptsize{Maltose utilization}\\
&  &  &  & \scriptsize{malK-lamB-malM} & \\
&  &  &  & \scriptsize{malS} & \\
\hline
& \scriptsize{16} & \scriptsize{crp} & \scriptsize{araC} & \scriptsize{araBAD} & \scriptsize{Arabinose utilization}\\
&  &  &  & \scriptsize{araE} & \\
&  &  &  & \scriptsize{araFGH} & \\
&  &  &  & \scriptsize{araJ} & \\
\scriptsize{4} & \scriptsize{17} & \scriptsize{rob} & \scriptsize{marRAB} & \scriptsize{fumC} & \scriptsize{Drug resistance}\\
&  &  &  & \scriptsize{nfo} & \\
&  &  &  & \scriptsize{sodA}&\\
&  &  &  & \scriptsize{zwf} & \\
\hline
\scriptsize{5}& \scriptsize{18} & \scriptsize{flhDC} & \scriptsize{fliAZY} & \scriptsize{flgBCDEFGHIJK} & \scriptsize{Flagella system}\\
&  &  &  & \scriptsize{flhBAE} & \\
&  &  &  & \scriptsize{fliE}&\\
&  &  &  & \scriptsize{fliFGHIJK} & \\
&  &  &  & \scriptsize{fliLMNOPQR} & \\
\hline
\scriptsize{7}& \scriptsize{19} & \scriptsize{fnr} & \scriptsize{arcA} & \scriptsize{cydAB} & \scriptsize{Anaerobic metabolism}\\
&  &  &  & \scriptsize{cyoABCDE} & \\
&  &  &  & \scriptsize{focA-pflB}&\\
&  &  &  & \scriptsize{glpACB} & \\
&  &  &  & \scriptsize{icdA} & \\
&  &  &  & \scriptsize{nuoABCDEFGHIJKLMN} & \\
&  &  &  & \scriptsize{sdhCDAB-b0725-sucABCD} &\\
\hline

\end{tabular}
\caption{\textbf{Feedforward loops in \emph{E. coli} transcription
network classified into multi-Z complexes.} Complex size is the
number of operons (Z-role nodes) in the FFL generalization}
\label{Table2}
\end{table*}

\begin{table*}
\begin{tabular}{|c|l|l|l|l|l|}
\hline
\scriptsize{Complex} & \scriptsize{Id}. & \scriptsize{X} & \scriptsize{Y} & \scriptsize{Z} & \scriptsize{Function}\\
\scriptsize{size} &  &  &  &  & \\
\hline
& \scriptsize{1} & \scriptsize{TUP1} & \scriptsize{RME1} & \scriptsize{IME1} & \scriptsize{Meiosis}\\
& \scriptsize{2} & \scriptsize{RIM101} & \scriptsize{IME1} & \scriptsize{DIT1} & \scriptsize{Sporulation}\\
& \scriptsize{3} & \scriptsize{MIG1} & \scriptsize{HAP2-3-4-5} & \scriptsize{CYC1} & \scriptsize{Formation of apocytochromes}\\
& \scriptsize{4} & \scriptsize{MIG1} & \scriptsize{GAL4} & \scriptsize{GAL1} & \scriptsize{Galactokinase}\\
\scriptsize{1}  & \scriptsize{5} & \scriptsize{MIG1} & \scriptsize{CAT8} & \scriptsize{JEN1}  & \scriptsize{Lactate uptake}\\
& \scriptsize{6} & \scriptsize{MIG2} & \scriptsize{CAT8} & \scriptsize{JEN1} & \scriptsize{(2X-FFL complex)}\\
& \scriptsize{7} & \scriptsize{GAT1} & \scriptsize{DAL80-GZF3} & \scriptsize{GAP1} & \scriptsize{Nitrogen utilization}\\
& \scriptsize{8} & \scriptsize{TUP1} & \scriptsize{ALPHA1} & \scriptsize{MFALPHA1} & \scriptsize{Mating factor alpha}\\
& \scriptsize{9} & \scriptsize{GAL11} & \scriptsize{ALPHA1} & \scriptsize{MFALPHA1} & \scriptsize{(2X-FFL complex)}\\
\hline
& \scriptsize{10} & \scriptsize{TUP1} & \scriptsize{ROX1} & \scriptsize{ANB1} & \scriptsize{Anaerobic metabolism}\\
&  &  &  & \scriptsize{CYC7} & \\
& \scriptsize{11} & \scriptsize{GLN3} & \scriptsize{GAT1} & \scriptsize{GAP1} & \scriptsize{Nitrogen utilization}\\
&  &  &  & \scriptsize{GLN1} & \scriptsize{Glutamate synthetase}\\
\scriptsize{2} & \scriptsize{12} & \scriptsize{GLN3} & \scriptsize{GAT1} & \scriptsize{DAL80} & \scriptsize{Nitrogen utilization}\\
&  &  &  & \scriptsize{GLN1} & \scriptsize{Glutamate synthetase}\\
& \scriptsize{13} & \scriptsize{GLN3} & \scriptsize{DAL80-GZF3} & \scriptsize{GAP1} & \scriptsize{Nitrogen utilization}\\
&  &  &  & \scriptsize{UGA4} & \\
& \scriptsize{14} & \scriptsize{PDR1} & \scriptsize{YRR1} & \scriptsize{SNQ2} & \scriptsize{Drug resistance}\\
&  &  &  & \scriptsize{YOR1} & \\
& \scriptsize{15} & \scriptsize{GCN4} & \scriptsize{MET4} & \scriptsize{MET16} & \scriptsize{Methionine biosynthesis}\\
&  &  &  & \scriptsize{MET17} & \\

\hline
 & \scriptsize{16} & \scriptsize{HAP1} & \scriptsize{ROX1} & \scriptsize{ERG11} & \scriptsize{Anaerobic metabolism}\\
&  &  &  & \scriptsize{HEM13} & \\
\scriptsize{3}&  &  &  & \scriptsize{CYC7} & \\
\hline
& \scriptsize{17} & \scriptsize{SPT16} & \scriptsize{SWI4-SWI6} & \scriptsize{CLN1} & \scriptsize{Cell cycle and}\\
&  &  &  & \scriptsize{CLN2} & \scriptsize{mating type switch}\\
&  &  &  & \scriptsize{HO} & \\
\hline

& \scriptsize{18} & \scriptsize{GCN4} & \scriptsize{LEU3} & \scriptsize{ILV1} & \scriptsize{Leucine and branched amino}\\
&  &  &  & \scriptsize{ILV2} &  \scriptsize{acid biosynthesis}\\
&  &  &  & \scriptsize{ILV5} & \\
\scriptsize{4}&  &  &  & \scriptsize{LEU4} & \\
 & \scriptsize{19} & \scriptsize{UME6} & \scriptsize{INO2-INO4} & \scriptsize{CHO1} & \scriptsize{Phospholipid biosynthesis}\\
&  &  &  & \scriptsize{CHO2} & \\
&  &  &  & \scriptsize{INO1}&\\
&  &  &  & \scriptsize{OPI3} & \\
\hline
\scriptsize{6}& \scriptsize{20} & \scriptsize{PDR1} & \scriptsize{PDR3} & \scriptsize{HXT11} & \scriptsize{Drug resistance}\\
&  &  &  & \scriptsize{HXT9} & \\
&  &  &  & \scriptsize{IPT1}&\\
&  &  &  & \scriptsize{PDR5} & \\
&  &  &  & \scriptsize{SNQ2} & \\
&  &  &  & \scriptsize{YOR1} & \\
\hline
\scriptsize{15}& \scriptsize{21} & \scriptsize{GLN3} & \scriptsize{DAL80} & \scriptsize{CAN1} & \scriptsize{Nitrogen utilization}\\
&  &  &  & \scriptsize{DAL1} & \\
&  &  &  & \scriptsize{DAL2}&\\
&  &  &  & \scriptsize{DAL3} & \\
&  &  &  & \scriptsize{DAL4} & \\
&  &  &  & \scriptsize{DAL5} & \\
&  &  &  & \scriptsize{DAL7} & \\
&  &  &  & \scriptsize{DCG1} & \\
&  &  &  & \scriptsize{DUR1} & \\
&  &  &  & \scriptsize{DUR3} & \\
&  &  &  & \scriptsize{GDH1} & \\
&  &  &  & \scriptsize{PUT1} & \\
&  &  &  & \scriptsize{PUT2} & \\
&  &  &  & \scriptsize{PUT4} & \\
&  &  &  & \scriptsize{UGA1} & \\
\hline

\end{tabular}
\caption{\textbf{Feedforward loops in \emph{S. cerevisiae}
transcription network classified into multi-Z complexes.} Complex
size is the number of genes (Z-role nodes) in the FFL
generalization.} \label{Table3}
\end{table*}

In order to compute the statistical significance of a certain
generalization of a motif $S$, we first find for each appearance
of $S$ in the network the maximal size generalization in which it
appears. Then we count the cumulative number of times $S$ appears
in the union of all the maximal generalizations (up to size k). In
order to verify that the generalization significance is not due to
many stand-alone appearances of the basic subgraph (e.g. a
single-Z FFL in the case of multi-Z FFL generalization), we
subtract the number of time $S$ appears as a stand-alone structure
in the network from the cumulative results (Note that in Fig 3 we
show the results before subtractions). We compare these numbers to
the corresponding numbers in randomized networks (Here we used
$Zscore > 2$). It is important to note that the randomized
networks preserve the incoming, outgoing and mutual edge degree
for each node. The networks are not constrained to have the same
number of 3-node or higher subgraphs as in the real network (in
~\cite{Milo} in contrast, 4-node motifs were detected based on
randomized networks that
preserved 3-node subgraph counts).\\

The network is described by a directed interaction graph
$G=(V,E)$, where $V$ is the set of nodes and $E$ is the set of
edges. An edge $(v_{i},v_{j}) \in {E} $ represents a directed link
between nodes $v_{i}$ and $v_{j}$. For every n-node subgraph S
which is detected as a network motif ~\cite{Shen-Orr,Milo} we
search for its simple generalizations (multiplicity of one of the
roles). We begin by building an induced graph $G'=(V',E')$. The
nodes in $G'$ are only those that act as members (nodes) of S
appearances in $G$, and the edges are only the edges in $G$
between these nodes. $G'$ is usually a much smaller graph then
$G$, but it contains all the information we need for our purpose.
For each simple generalization type j (multiplicity of the j-th
role of the subgraph) the following is performed: A non-directed
graph $\hat{G}=(\hat{V},\hat{E})$ is built where each node
represents a specific basic subgraph $S$ in $G$ (a specific set of
nodes in $G$ that form a subgraph of type $S$). The number of
nodes in $\hat{G}$ equals the number of times $S$ appears in the
original graph $G$. Two nodes in $\hat{G}$ are connected if and
only if they follow the generalization type, j, and the
generalization rule (strong or weak). Setting the edges in
$\hat{G}$ is done efficiently by using the appearances of the
basic subgraph in $G'$ as starting points. For each specific
'starting point' subgraph $S_{1}$ in $G'$ we pass through all the
'neighboring' subgraphs $S_{2}$ ('neighboring' in the sense that
they share all node roles excluding j-th node roles) and check if
the joint subgraph $(S_{1}\bigcup S_{2})$ in $G'$ forms a
generalization type j. After setting all edges in $\hat{G}$, the
next step is to find all maximal cliques ~\cite{Bron} (a group of
nodes in which every two are connected) in $\hat{G}$. Each maximal
clique represents a maximal generalization type j of $S$ (i.e. the
generalization with maximal number of appearances of the basic
subgraph). We store the size and the members (nodes in the
original network) of all maximal generalizations. Complex
generalizations (when more than one role is replicated) were
detected in a similar way by
appropriately changing the rules for setting the edges in $\hat{G}$.\\

\section{Network databases}
Transcription network of \emph{E.coli} ~\cite{Shen-Orr}, version
1.1 (N=423, E=519) available at
http://www.weizmann.ac.il/mcb/UriAlon/ was based on selected data
from ~\cite{Milo,Salgado} and literature. Transcription network of
yeast (\emph{S. cerevisiae}) ~\cite{Milo}, version 1.3 (N=685,
E=1052) available at http://www.weizmann.ac.il/mcb/UriAlon/ was
based on selected data from ~\cite{Milo,Costanzo}. (N=number of
nodes, E=number of edges). Self edges were excluded. The Neuronal
synaptic connection network of \emph{C. elegans}  (N=280, E=400)
was based on ~\cite{White} as arranged in ~\cite{AY}. The network
was compiled with a cutoff of at least 5 synapses for connections
between neurons. Target muscle cells were excluded. Electronic
forward-logic chips ~\cite{Milo} were obtained by parsing the
ISCAS89 benchmark data set ~\cite{Brglez} available at
www.cbl.ncsu.edu/CBL\_Docs/iscas89.html . Bi-fan generalizations
data (Table 1) are shown for chip S15850 (N=10383, E=14240), and
are representative of all logic chips in the database.


\newpage

\begin{thebibliography}{99}

\bibitem{Hartwell}
L.H. Hartwell, J.J. Hopfield, S. Leibler, and A.W. Murray,
\textrm{Nature} \textbf{402}: C47-52 (1999).

\bibitem{Ouzounis}
C.A. Ouzounis and P.D. Karp, \textrm{Genome Res.} \textbf{10}:
568-576 (2000).

\bibitem{McAdams2000}
H. McAdams and A. Arkin, \textrm{Curr. Biol.} \textbf{10}:
R318-320 (2000).

\bibitem{Elowitz}
M.B. Elowitz,  and S. Leibler, \textrm{Nature}
\textbf{403}:335-338 (2000).

\bibitem{Savageau}
M.A. Savageau, \textrm{Chaos} \textbf{11}: 142-159 (2001).

\bibitem{Rao}
C.V. Rao and A.P. Arkin, \textrm{Annu. Rev. Biomed. Eng.}
\textbf{3}: 391-419 (2001).

\bibitem{Strogatz}
S.H. Strogatz,  \textrm{Nature} \textbf{410}: 268-276 (2001).

\bibitem{Bolouri}
H. Bolouri and E.H. Davidson \textrm{Bioessays}
\textbf{24}:1118-1129 (2002).

\bibitem{Hasty}
J. Hasty, D. McMillen, and J.J. Collins \textrm{Nature}
\textbf{420}: 224-230 (2002).

\bibitem{Guet}
C.C. Guet, M.B. Elowitz, W. Hsing, and S. Leibler,
\textrm{Science} \textbf{296}: 1466-1470 (2002).

\bibitem{Tyson}
J.J. Tyson, K.C. Chen, and B. Novak, \textrm{Curr. Opin. Cell
Biol.} \textbf{15}: 221-231 (2003).

\bibitem{Sneppen}
S. Maslov. K. Sneppen, \textrm{Science} \textbf{296}(5569):910-3
(2002).

\bibitem{NewmanReview}
M. Newman, \textrm{SIAM Review} \textbf{45}: 167-256 (2003).

\bibitem{Milo}
R. Milo, S. Shen-Orr, S. Itzkovitz, N. Kashtan, D. Chklovskii, and
U. Alon, \textrm{Science} \textbf{298}: 824-827 (2002).

\bibitem{Shen-Orr}
S. Shen-Orr, R. Milo, S. Mangan, and U. Alon, \textrm{Nat. Genet.}
\textbf{31}: 64-68 (2002).

\bibitem{Milo_2004}
R. Milo et. al.  \textrm{Science} \textbf{303}: 1538-1542 (2004).

\bibitem{Lee}
T.I. Lee et. al. \textrm{Science} \textbf{298}: 799-804 (2002).

\bibitem{Mangan}
S. Mangan and U. Alon, \textrm{Proc. Natl. Acad. Sci. U S A.}
\textbf{100}(21):11980-5 (2003)

\bibitem{Mangan_JMB}
S. Mangan, A. Zaslaver, and U. Alon, \textrm{J. Mol. Biol.}
\textbf{334}(2):197-204 (2003).

\bibitem{Barabasi_BMC}
R. Dobrin, Q.K.Beg, A.L. Barabasi, Z.N. Oltvai, \textrm{BMC
Bioinformatics} \textbf{5}(1):10 (2004).

\bibitem{Ronen}
M. Ronen, R. Rosenberg, B.I. Shraiman, and U. Alon, \textrm{Proc.
Natl. Acad. Sci. U S A.} \textbf{99}: 10555-10560 (2002).

\bibitem{Zaslaver}
A. Zaslaver, A. Mayo, M. Surette, N. Rosenberg, P. Bashkin, H.
Sberro,  M. Tsalyuk, and U. Alon, \textrm{Nat. Genet.}
\textbf{36}(5): 486-91 (2004).

\bibitem{Yuh}
C.H. Yuh, H. Bolouri, and E.H. Davidson \textrm{Science}
\textbf{279}: 1896-1902 (1998).

\bibitem{Hwa}
N. Buchler, U. Gerland, and T. Hwa, \textrm{Proc. Natl. Acad. Sci.
U S A.} \textbf{100}: 5136-5141 (2003).

\bibitem{Setty}
Y. Setty, A.E. Mayo, M.G. Surette, and U. Alon, \textrm{Proc.
Natl. Acad. Sci. U S A.} \textbf{100}(13):7702-7 (2003).

\bibitem{Itzkovitz}
S. Itzkovitz, R. Milo, N. Kashtan, G. Ziv, and U. Alon. Phys Rev.
E \textbf{68}: 026127 (2003).

\bibitem{Kashtan}
N. Kashtan, S. Itzkovitz, R. Milo, and U. Alon,
\textrm{Bioinformatics}, advance access
'10.1093/bioinformatics/bth163' (2004).

\bibitem{Nesetril}
J. Nesetril and S. Poljak, \textrm{Commen. Math. Univ. Carol.}
\textbf{26}: 415-419 (1985).

\bibitem{Harary}
F. Harary,  and E.M. Palmer, \textit{Graphical Enumeration}.
(Academic Press, NY, 1973).

\bibitem{White}
J. White, E. Southgate, J. Thomson, and S. Brenner, P\textrm{hil.
Trans. Roy. Soc. London Ser. B} \textbf{314}: 1-340 (1986).

\bibitem{AY}
T.B. Achacoso and W.S. Yamamoto, \textit{AY's Neuroanatomy of
C.elegans for Computation}. (CRC Press, 1992).

\bibitem{Neurons_Comment}
We note that in the neuronal network where edges represent all
synaptic connections (not only those with 5 or more synapses), we
find numerous examples of the multi-Z and multi-Y FFLs, with the
multi-X FFL the most common structure (data not shown).

\bibitem{Brglez}
F. Brglez, D. Bryan, and K. Kozminski \textit{Combinational
Profiles of Sequential Benchmark Circuits. Proc}. IEEE Int.
Symposium on Circuits and Systems: 1929-1934 (1989).

\bibitem{Sole}
R.F. Cancho, C. Janssen, and R. V. Solé \textrm{Phys. Rev. E}
\textbf{64}: 046119 (2001).

\bibitem{Weiss}
S. Basu, R. Mehreja, S. Thiberge, M.T. Chen, R. Weiss,
\textrm{Proc Natl Acad Sci U S A.} \textbf{101}(17):6355-60
(2004).

\bibitem{McAdams1998}
H. McAdams and A. Arkin, \textrm{Annu. Rev. Biophys. Biomol.
Struct.} \textbf{27}: 199-224 (1998).

\bibitem{Rosenfeld}
N. Rosenfeld, M.B. Elowitz, and U. Alon, \textrm{J. Mol. Biol.}
\textbf{323}: 785-793 (2002).

\bibitem{Rosenfeld_2003}
N. Rosenfeld and U. alon, \textrm{J. Mol. Biol.}
\textbf{329}:645-654 (2003).

\bibitem{Kalir}
S. Kalir, J. McClure, K. Pabbaraju, C. Southward, M. Ronen, s.
Leibler, M.G. Surette, U. Alon, \textrm{Science} \textbf{292}:
2080-2083 (2001).

\bibitem{Kalir_Cell}
S. Kalir and U. Alon . \textrm{Cell} in press (2004).

\bibitem{Chalfie}
M. Chalfie, J.E. Sulston, J.G. White, E. Southgate, J.N. Thomson,
and S. Brenner, \textrm{The Journal of Neuroscience} \textbf{5}:
956-964 (1985).

\bibitem{Hope}
I.A. Hope \textit{C. elegans A practical approach}. (Exford
university press, 1999).

\bibitem{Goodman}
M.B. Goodman, D.H. Hall, L. Avery, and S.R. Lockery,
\textrm{Neuron} \textbf{20}: 763-772 (1998).

\bibitem{Hansen}
M.C. Hansen, H. Yaclin, and J.P. Hayes, \textit{Unveiling the
ISCAS-85 Benchmarks: A case study in reverse engineering}. IEEE
Design and Test: 72-80 (1999).

\bibitem{Conant}
G.C. Conant, and A. Wagner, \textrm{Nat. Genet.} \textbf{34}:
264-266 (2003).

\bibitem{Wasserman}
S. Wasserman and K. Faust, . \textit{Social Network Analysis}.
(Cambridge University Press, Cambridge, 1994).

\bibitem{Lorrain}
F. Lorrain and H.C.White, \textrm{Journal of Mathematical
Sociology} \textbf{1}: 49-80 (1971).

\bibitem{Winship}
C. Winship. \textrm{Social Networks} \textbf{10}: 209-231 (1988).

\bibitem{Winship&Mandel}
C. Winship and M. Mandel, \textrm{Sociological Methodology}
1983-1984: 314-344 (1983).

\bibitem{Everett}
M.G. Everett, J.P. Boyd, and S.P. Borgatti, \textrm{Journal of
Mathematical Sociology} \textbf{15}: 163-172 (1990).

\bibitem{Bron}
C. Bron, J. Kerbosch, \textrm{Commun. ACM.} \textbf{16}: 575-577
(1973).

\bibitem{Salgado}
H. Salgado, A. Santos-Zavaleta, S. Gama-Castro, D. Millan-Zarate,
E. Diaz-Peredo, F. Sanchez-Solano, E. Perez-Rueda, C.
Bonavides-Martinez, and J. Collado-Vides, \textrm{Nucleic Acids
Res.} \textbf{29}: 72-74 (2001).

\bibitem{Costanzo}
M.C. Costanzo et. al. \textrm{Nucleic Acids Res.} \textbf{29}:
75-79 (2001).











\end{thebibliography}
\end{document}